\documentclass[onecolumn, aps, prd, groupadress, nofootinbib, superscriptaddress, showkeys, showpacs]{revtex4-1}

\usepackage{amsmath, amsfonts, amssymb, mathrsfs}
\usepackage{amsthm}
\usepackage{tensor}
\usepackage{multirow}
\usepackage{bbm} 
\usepackage[geometry]{ifsym}

\usepackage{cmbright}

\usepackage{graphicx}
\usepackage{float}
\usepackage[font=small]{caption}

\usepackage[hyperindex]{hyperref}
\hypersetup{breaklinks=true, hyperfootnotes=true, pagecolor=white, colorlinks=true}
\usepackage[all]{hypcap}

\makeatletter
\DeclareRobustCommand*{\bfseries}{%
  \not@math@alphabet\bfseries\mathbf
  \fontseries\bfdefault\selectfont
  \boldmath
}
\makeatother

\renewcommand{\nabla}{\!\mathrel{\raisebox{.15em}{%
           \reflectbox{\rotatebox[origin=c]{180}{$\triangle$}}}}\!\!} 

\newcommand{\Dslash}{{D\kern-0.63em{/}}}

\begin{document}

\title{The cosmological constant and a scalar field coupled non minimally to gravity}
\

\author { M. Novello and A.E.S. Hartmann}
 \affiliation{
Centro de Estudos Avan\c{c}ados de Cosmologia/CBPF \\
Rua Dr. Xavier Sigaud 150, Urca 22290-180 Rio de Janeiro, RJ-Brazil}
\date{\today}

\vspace{2 cm}

\date{\today}

\begin{abstract}
We show that the combined minimal and non minimal interaction with the gravitational field may produce the generation of a cosmological constant without self-interaction of the scalar field. In the same vein we analyze the existence of states of a scalar field that by a combined interaction of minimal and non minimal coupling with the gravitational field can exhibit an unexpected property, to wit, they are acted on by the gravitational field but do not generate gravitational field. In other words, states that seems to violate the action-reaction principle. We present explicit examples of this situation in the framework of a spatially isotropic and homogeneous universe.
\end{abstract}

\maketitle

\section{Introduction}

The distribution of the energy of a scalar field dependent only on time in a space time endowed with a spatially isotropic and homogeneous metric may be represented as a perfect fluid even in the case of a non minimal coupling to gravity. This allows a very unexpected result to appear, that is the possibility of the scalar field to be acted upon by the gravitational field that does not drive the gravitational field, as a sort of violation of the action-reaction principle,in the realm of General Relativity theory. We will show here how this is possible.

From the quantum theory of fields the mechanism of spontaneous symmetry breaking set in evidence a sort of dynamical origin of a cosmological constant. In a sense, this opened the way to allow the proposed scenario of inflation \cite{guth}, \cite{starobinski}, \cite{novellosalim}. The idea is very simple. Start with a scalar field the dynamics of which, besides its kinematical term is provided by a potential $ V(\varphi)$ that is a polynomial. A typical case, transformed in a paradigm, is $V = c_{1} \, \varphi^{2} + c_{2} \, \varphi^{4}$ for a convenient choice of constants $c_{1}$ and $ c_{2}.$ The density of energy is associated to the constant $\varphi_{0}$ that extremizes $ V.$ A further restriction on the constants makes this extremum a stable one for the case $ \varphi_{0}.$ Then this constant field shows a non trivial energy distribution having the appearance of a cosmological constant.

In the present paper we show an alternative of associating the energy distribution of the scalar field (without self-interaction) to a cosmological constant related to a combined interaction of gravity to the scalar field.

\section{Dynamics: Minimal coupling}
We start by making a short resume of properties that will be used in our text. In the framework of standard cosmology the geometry of the universe is given by a spatially isotropic and homogeneous metric with Euclidean section

\begin{equation}
ds^{2} = dt^{2} - a^{2}(t) \, (dx^{2} + dy^{2} + dz^{2})
\end{equation}
In this case
$$ R_{00} = \dot{\theta} + \frac{1}{3} \, \theta^{2}, $$
$$ R_{ij} = \frac{1}{3} \,( \dot{\theta} +  \, \theta^{2}) \, g_{ij} $$
$$ R = 2 \, \dot{\theta} +  \frac{4}{3}\, \, \theta^{2}$$ where we have defined the expansion factor $ \theta = 3 \, \dot{a}/a$ and a dot means time derivative.

Let us first examine the simplest case of the Lagrangian of a massless scalar field

\begin{equation}
L = \frac{1}{2 \kappa} \, R + \frac{1}{2} \, \partial_{\mu} \varphi \, \partial^{\mu} \varphi - \frac{\Lambda}{\kappa} \label{12jan1}
\end{equation}
The equation of motion of the scalar field and the metric are given by:

\begin{equation}
\Box \,\varphi = 0
\label{12 jan 5}
\end{equation}

\begin{equation}
R_{\mu\nu} - \frac{1}{2} \, R \, g_{\mu\nu} = - \, \kappa \, T_{\mu\nu}^{0} - \Lambda \, g_{\mu\nu}
\label{12jan2}
\end{equation}
where (note that we use indiscriminately $ \varphi_{\mu} \equiv \varphi_{, \, \mu}$)

\begin{equation}
 T_{\mu\nu}^{0} = \varphi_{, \, \mu} \, \varphi_{, \, \nu}  - \frac{1}{2} \, \varphi_{\alpha} \, \varphi^{\alpha} \, g_{\mu\nu}.
 \label{14jan6}
 \end{equation}

The unique non trivial equations reduces to this case to the conservation of energy and the evolution of the expansion factor, that is

\begin{equation}
 \dot{\rho} + ( \rho + p ) \, \theta = 0.
 \label{12jan3}
 \end{equation}

 \begin{equation}
 \rho + \Lambda= \frac{1}{3} \, \theta^{2}
  \label{12jan4}
 \end{equation}

 where, in our case
 $$ \rho = p = \frac{1}{2} \, \dot{\varphi}^{2}. $$
 It then follows that the density of energy
 $$ \rho = \rho_{0} \, a^{-6}.  $$
 Thus, for very small values of $ a(t)$ the cosmological constant can be completely neglected. In fact, comparison of the energy of the scalar field to others forms of matter, like the energy of an electromagnetic fluid for instance, where the density of the photons is provided by  $ \rho_{\gamma} = constant \, a^{-4},$ shows that such scalar field dominates the cosmical scenario in the primordial era where the scalar function $ a(t)$ is very small.

 In this case the solution of these equations are immediate

 $$  \varphi = \left(\frac{2}{3}\right)^{1/2} \, \ln t  $$
 $$ a(t) = a_{0} \, t^{1/3}.$$

 Let us now introduce a non minimal coupling and examine its consequences. We shall see that in certain cases, the combined coupling minimal plus non minimal can supress totally the effect of the scalar field on the geometry.

\section{Non minimal coupling: suppressing the energy}

Let us set the Lagrangian

\begin{equation}
L = \frac{1}{2 \kappa} \, R + \frac{1}{2} \, \partial_{\mu} \varphi \, \partial^{\mu} \varphi + B(\varphi)
\, R - \frac{1}{\kappa} \, \Lambda \label{210}
\end{equation}
where $ B(\varphi)$ is for the time being an arbitrary function of the scalar field. Then, it yields the equation for the metric as

\begin{equation}
  R_{\mu\nu} - \frac{1}{2} \, R \, g_{\mu\nu} = - \, T_{\mu\nu}^{0} - \,  Z_{\mu\nu} - \Lambda \, g_{\mu\nu}
  \label{gr1}
  \end{equation}
where $ T_{\mu\nu}^{0} $ is given by (\ref{14jan6}) and
\begin{equation}
Z_{\mu\nu} = 2 B \, ( R_{\mu\nu} - \frac{1}{2} \, R \, g_{\mu\nu} ) + 2 \nabla_{\mu} \nabla_{\nu} B - 2 \Box  B \, g_{\mu\nu}
\end{equation}

In the case the scalar field depends only on the cosmological time we obtain for the density of energy the forms
$$ p_{0} = \rho_{0} = \frac{1}{2} \, \dot{\varphi}^{2}$$
and for the non minimal part

$$ \rho_{Z} = - \, \frac{2}{3} \, B \, \theta^{2} - 2 \, \dot{B} \, \theta. $$

$$p_{Z} = 2 \, \ddot{B} + \frac{4}{3} \, \dot{B} \, \theta +  \, \frac{2}{3} \, B \, \theta^{2} + \frac{4}{3} \, B \, \dot{\theta}  $$

We are looking for solution of the scalar field such that its total energy $ \rho = \rho_{0} +  \rho_{Z} $ and pressure $ p = p_{0} + p_{Z} $ vanishes, although its field satisfies the equation of motion of the curved space-time. The responsible for the curvature is the cosmological constant that generates a constant expansion factor $ \theta_{0}.$ Note that this situation can be generalized for other curved space-times.

We start by solving the equation of motion for $ \varphi.$ The equation is

 \begin{equation}
 \Box \, \varphi - R \, B^{'} = 0
 \label{7nan192}
 \end{equation}
where $ B^{'} $ represents derivative of $ B $ with respect to $ \varphi.$ Let us choose, just to present an specific case the non minimal term as given by
$$ B = b \, \varphi^{2}$$
and the equation of motion (in the case the field depends only on time) reduces to

\begin{equation}
\ddot{\varphi} + \theta_{0} \, \dot{\varphi} -  b \,  \frac{8}{3} \, \theta^{2}_{0} \, \varphi = 0.
\label{7jan193}
\end{equation}

We set a solution of the form
$$ \varphi =\varphi_{0} \, e^{m \, t} $$
The equation of motion is satisfied if $m$ and $ b$ are related by

\begin{equation}
m^{2} +  \theta_{0} \, m - \frac{8}{3} \, b \, \theta_{0}^{2} = 0.
\label{11jan1}
\end{equation}

Using this proposal into the two independent conditions
$$ \rho_{0} + \rho_{Z} = 0 $$
$$ p_{0} + p_{Z} = 0 $$
yields, respectively,  the equations:

\begin{equation}
m^{2} -  \frac{4}{3} b \, \theta_{0}^{2} - 8 \, b \, m \, \theta_{0} = 0.
\label{11jan2}
\end{equation}

\begin{equation}
m^{2} + 16 \, b \, m^{2} + \frac{4}{3} b \, \theta_{0}^{2} + \frac{16}{3} \, b \, m \, \theta_{0}  = 0.
\label{11jan3-}
\end{equation}

The solution of these equations gives the value  $ m = -  \, \sqrt{3 \, \Lambda}/3 = - \, 1/3 \, \theta_{0},$ and the value of the constant $ b = - \, 1/12.$ This means that the non minimal term is nothing but a conformal coupling.

\vspace{0.3cm}
We can then state the conclusion: \emph{The combined coupling of a minimal and conformal coupling of a massless scalar field to gravity can annihilate the effect of the scalar field on gravity in an universe of constant expansion.}

\subsection{The case of a variable expansion factor}

The suppressed process of the energy by non minimal gravitational interaction can be generalized for the case the expansion factor is not a constant. We set the Lagrangian

\begin{equation}
L = \frac{1}{2 \kappa} \, R + \frac{1}{2} \, \partial_{\mu} \varphi \, \partial^{\mu} \varphi + b \, \varphi^{2} \, R + V(\varphi) + L_{m}
\label{24jan1}
\end{equation}

and consider the matter term as generating an expanding universe with expansion factor $ \theta =\theta_{0}/t.$

The equation of motion of the scalar field is

$$ \Box \varphi - 2 \, b \, \varphi \, R - \frac{\delta V}{\delta \varphi}= 0 $$

Let us consider the case in which $ V = \lambda \, \varphi^{4}.$
Then, a solution is given by
$$ \varphi = \frac{\varphi_{0}}{t}.  $$

In the case of a conformal interaction $ b = - 1/12$ it follows that

$$ \lambda \,\varphi_{0}^{2} = \frac{1}{2} \, ( 1 - \frac{\theta_{0}}{3})^{2}.$$

It is then immediate to show that for the density and pressure of the scalar field we obtain

$$  \rho(\varphi) = p(\varphi) = 0.$$

That is, the non minimal coupling term annihilates the free part of the energy-momentum tensor of the scalar field.

\subsection{Second example}

Let us now consider the case that there is an electromagnetic field and set the Lagrangian as given by

\begin{equation}
L = \frac{1}{2 \kappa} \, R + \frac{1}{2} \, \partial_{\mu} \varphi \, \partial^{\mu} \varphi + B(\varphi)
\, R - \frac{1}{4} \, F_{\mu\nu} \, F^{\mu\nu} \label{13jan1}
\end{equation}

We are interested in analyze the effects of the non minimal coupling in the standard spatially isotropic and homogeneous metric. To be consistent with the symmetries of this choice of the metric,
an averaging procedure must be performed if electromagnetic fields
are to be taken as the source of the gravitational field, according to the standard procedure \cite{tolmanbook}. As a consequence, the components of the electric
$E_{i}$ and magnetic $H_{i}$ fields must satisfy the following
relations:
\begin{eqnarray}
<{E}_i > = 0,\qquad
%
<{H}_i> &=& 0,\qquad
%
<{E}_i\, {H}_j> = 0,
\label{meanEH}\\[1ex]
<{E}_i\,{E}_j> &=& -\, \frac{1}{3} {E}^2
\,g_{ij},
\label{meanE2}\\[1ex]
<{H}_i\, {H}_j> &=&  -\, \frac{1}{3} {H}^2
\,g_{ij}.
\label{meanH2}
\end{eqnarray}
where $E$ and $H$ depends only on time.

Using the above average values it follows that the Maxwell energy-momentum tensor $T_{\mu\nu}^{m}$
reduces to a perfect fluid configuration with energy density
$\rho_\gamma$ and pressure $p_\gamma$ given by
\begin{equation}
<T_{\mu\nu}^{\gamma} > = (\rho_\gamma + p_\gamma)\,
v_{\mu}\, v_{\nu} - p_\gamma\, g_{\mu\nu}, \label{Pfluid}
\end{equation}
where
\begin{equation}
\label{RhoMaxwell} \rho_\gamma = 3p_\gamma = \frac{1}{2}\,(E^2 +
{H}^2).
\end{equation}
From the conservation of the energy-momentum tensor (note that there is no direct interaction between the scalar and the electromagnetic fields) it follows that the density of energy goes like $ a^{-4}$.
Consequently we obtain the standard value for the scale factor once, as we shall see, there is no contribution coming from the scalar field to change the curvature of space-time
$$ a(t) = a_{0} \, \sqrt{t}$$

The equation of motion for the scalar field reduces to
$$ \Box \, \varphi = 0$$ that is
$$ \varphi = \frac{\varphi_{0}}{\sqrt{t}}$$
Let us set
$$ B = b \, \varphi^{2}$$
Then we obtain for the annihilation of the density of energy of the scalar field the two conditions

$$\rho_{0} +  \rho_{Z} = \frac{1}{2} \, \dot{\varphi}^{2} - \, \frac{2}{3} \, B \, \theta^{2} - 2 \, \dot{B} \, \theta = 0. $$

$$ p_{0} +  p_{Z} = \frac{1}{2} \, \dot{\varphi}^{2} + 2 \, \ddot{B} + \frac{4}{3} \, \dot{B} \, \theta +  \, \frac{2}{3} \, B \, \theta^{2} + \frac{4}{3} \, B \, \dot{\theta} = 0. $$

These equations are satisfied only in the case $ b = - 1/12$ that is in the case the scalar field is coupled conformal to gravity.

\vspace{0.3cm}
We can then state the conclusion: \emph{The combination of a minimal and conformal coupling of a massless scalar field to gravity can annihilate the effect of the scalar field on gravity in the presence of a cosmological scenario driven by a gas of photons. }

\section{Curvature induced cosmological constant}

Let us now follow another path and show how it is possible that a combination of the minimal and a specific non minimal interaction of a scalar field to gravity can generate a cosmological constant. We start by a very short review of the original framework of inflationary universe \cite{linde}.

In order to explore the consequences of a primordial phase in which the universe has a constant expansion factor, $\theta = \theta_{0}$ in our notation, it was suggested to make appeal to the spontaneous symmetry breaking mechanism that allows a simple configuration of a constant scalar field controlled by the minimum of a given potential $ V(\varphi)$ in the case the Lagrangian takes the form

\begin{equation}
L = \frac{1}{2 \kappa} \, R + \frac{1}{2} \, \partial_{\mu} \varphi \, \partial^{\mu} \varphi + V(\varphi)
\label{21jan1}
\end{equation}
where we set $ \hbar = c = 1.$
In a spatially homogeneous and isotropic geometry the scalar field must depend only on the cosmical time. Then its energy density and pressure are given by

$$\rho = \frac{1}{2} \, \dot{\varphi}^{2} -  V $$

$$ p = \frac{1}{2} \, \dot{\varphi}^{2} + V. $$
The equation of motion takes the form

$$ \Box \varphi - \frac{\delta V}{\delta \varphi} = 0$$

For constant solution, in the extremum of the potential, one obtains trivially the condition that turns the distribution of the energy of the field into a cosmological constant. Thus the true origin of the cosmological constant in this framework rests on the self-interaction contained in the potential $ V(\varphi).$

Let us now show that there is another possibility that does not need such self-interaction but instead use the non minimal coupling of the scalar field to gravity to generate a cosmological constant.

The Lagrangian for a massive scalar field coupled non minimally to gravity is given by

\begin{equation}
L = \frac{1}{2 \kappa} \, R + \frac{1}{2} \, \partial_{\mu} \varphi \, \partial^{\mu} \varphi + b \, \varphi^{2} \, R - \frac{1}{2} \, \mu^{2} \, \varphi^{2}
\label{14jan5}
\end{equation}

In the framework of a spatially isotropic and homogeneous space time with Euclidean section driven by the cosmological constant (where the expansion factor is constant $ \theta_{0}$ ) the minimal coupling part of the energy-momentum of the field takes the form

$$ \rho_{0} =  \frac{1}{2} \, \dot{\varphi}^{2} + \frac{1}{2} \, \mu^{2} \, \varphi^{2}  $$

$$ p_{0} =   \frac{1}{2} \, \dot{\varphi}^{2}  - \frac{1}{2} \, \mu^{2} \, \varphi^{2}   $$

and for the non minimal coupling it yields

$$ \rho_{Z} =  - \, \frac{2}{3} \, B \, \theta^{2} - 2 \, \dot{B} \, \theta. $$

$$  p_{Z} =  2 \, \ddot{B} + \frac{4}{3} \, \dot{B} \, \theta +  \, \frac{2}{3} \, B \, \theta^{2} + \frac{4}{3} \, B \, \dot{\theta}  $$
in which $ B = b \, \varphi^{2}.$

The case of a constant field $ \varphi = \varphi_{0}$ produces an expanding universe (Euclidean section) with a constant expansion $ \theta = \theta_{0}.$   The equation of motion of the scalar field combined with the equation of the metric $ \kappa \,\rho = \theta^{2}/3c^{2}$ implies that
$$ \rho_{0} + \rho_{Z} + p_{0} + p_{Z} = 0 $$

$$ \theta_{0}^{2} = \frac{3}{8} \,  \frac{\mu^{2}}{b}$$

$$ \varphi_{0}^{2} = \frac{4}{3} \, \left( \frac{\theta_{0}}{\mu} \right)^{2}. $$
In other words, a cosmological constant appears thanks to the non minimal coupling given by

$$ \Lambda = \frac{1}{4} \,  \mu^{2} \, \varphi^{2}_{0}= \frac{1}{3} \, \theta_{0}^{2}.$$

This solution is stable. This can be easily shown by making the perturbation $\varphi_{0} \rightarrow \varphi_{0} + \delta \varphi$ and $ \theta_{0} \rightarrow \theta_{0} + \delta \theta.$ The time dependance is given by
$$ \delta \varphi = \delta \varphi_{i} \, \exp{(- \theta_{0} \, t)}.$$

Note that in the present case the non minimal coupling cannot be a conformal one \cite{com}.

\subsection{ Spontaneous symmetry breaking or non minimal coupling?}

Let us analyze now a generalization of both cases examined in the previous sections. We set for the Lagrangian

\begin{equation}
L = \frac{1}{2 \kappa} \, R + \frac{1}{2} \, \partial_{\mu} \varphi \, \partial^{\mu} \varphi +  V(\varphi) \left( 1 - \frac{2 \, b}{\mu^{2}} \, R \right)
\label{21jan5}
\end{equation}

where $ b $ is a dimensionless constant and for comparision to the standard framework we set for the potential the form

$$ V(\varphi) = - \, \frac{\mu^{2}}{2} \, \varphi^{2} + \lambda \, \varphi^{4}.$$

We limit the analysis to the cosmological scenario of spatially homogeneous and isotropic metric with Euclidean section for constant expansion where the scalar of curvature reduces to
$$ R = \frac{4}{3} \, \theta^{2}.$$

The equation of motion is given by

$$ \Box \varphi - \left( 1  - \frac{2 \, b}{\mu^{2}} \, R\right) \, \frac{\delta V}{\delta \varphi} = 0 $$

In the case of a constant solution $ \varphi = \varphi_{0}$ the energy-momentum of the field reduces to the density and pressure given by

$$ \rho =  - V + \frac{4}{3} \, \frac{b \, \theta^{2}}{\mu^{2}} \, V  $$
$$ p =   + V  - \frac{4}{3} \, \frac{b \, \theta^{2}}{\mu^{2}} \, V    $$
that is, the distribution of energy is nothing but an effective cosmological constant, that is, $ \rho + p = 0$

In this case, the dynamics of $\varphi$ admits two class of solutions:

\begin{equation}
\frac{\delta V}{\delta \varphi} = 0
\label{21jan6}
\end{equation}
or
\begin{equation}
 1  - \frac{2 \, b}{\mu^{2}} \, R  = 0
\label{21jan7}
\end{equation}

The first case led to the break of symmetry  scenario of standard inflation. The second one represents the combined effect of minimal and non minimal coupling of $\varphi$ field to gravity where the expansion factor is given by
$$ \theta_{0}^{2} = \frac{3}{8} \, \frac{\mu^{2}}{b}.$$

 In both cases the energy distribution of the scalar field is identical to a cosmological constant.

In the gravitational case the equation
$$ \rho = \frac{\theta^{2}}{3} $$
yields \cite{c1}
\begin{equation}
\varphi_{0}^{2} = \frac{\mu^{2}}{4 \, \lambda} \, \left( 1 + \sqrt{ 1 - \frac{32}{3} \, \frac{\lambda \, \theta^{2}_{0}}{\mu^{4}}} \right).
\label{21jan8}
\end{equation}

\section{Final comments}

We have presented some cases in which the effect of non minimal coupling of a scalar field  to the curvature of space-time allows the existence of states such that the net effect of the free part of the energy added to the gravitational interaction annihilate each other. This means that although the scalar field is embedded in a curved space its contribution to the metric vanishes. A similar interaction allows the generation of a cosmological constant without self interaction of the scalar field. In a forth coming work we examine these properties for other fields, in particular for the electromagnetic field \cite{novellohartmann}, \cite{hartmannenovello}.

\subsection{Acknowledgements}

We would like to acknowledge the financial support from brazilian agencies Finep, Faperj and CNPq.


\begin{thebibliography}{100}
\bibitem{com}  We note that the expression of the constant value of the field reads $ \varphi_{0}^{2} = 4\theta_{0}^{2}/3 \mu^{2} L_{Pl}^{2}.$ In the paper we are using units such that the Planck lenght is equal to $ 1.$
\bibitem{c1} Compare with the case of SSB in which the condition to extremize the potential yields the same result for the particular choice of constant $ \lambda $ given by $ \lambda =3 \, b \, \mu^{2}/4.$
\end{thebibliography}
\end{document}